\documentclass[reprint,
 amsmath,amssymb,
 aps, onecolumn
]{revtex4-2}
\usepackage{graphicx}
\usepackage{dcolumn}
\usepackage{bm}
\usepackage{hyperref}
\usepackage[mathlines]{lineno}
\usepackage{xcolor}

\begin{document}

\preprint{APS/123-QED}

\title{Rayleigh-Plateau Instability on an angled and eccentric wire}
\author{Dilip Kumar Maity$^1$}\email{dilip.maity@kaust.edu.sa} 
\author{Christopher Wagstaff$^2$} \author{Sandip Dighe$^1$} \author{Tadd Truscott$^1$}\email{taddtruscott@gmail.com}

\affiliation{ $^1$Department of Mechanical Engineering, Physical Science and Engineering Division, King Abdullah University of Science and Technology, Thuwal, 23955, Kingdom of Saudi Arabia. \\
$^2$Department of Chemical Engineering, King Abdullah University of Science and Technology, Thuwal, 23955, Kingdom of Saudi Arabia
}

\date{\today}

\begin{abstract}
This research explores the modulation of Rayleigh-Plateau instability by adjusting the orientation angle and eccentricity of a wire within a nozzle. We demonstrate that both the angle and eccentricity significantly influence the Rayleigh-Plateau instability regimes. They both also influence characteristics, such as bead velocity along the wire, bead spacing (wavelength), and bead volume. Notably, when wires are both angled and eccentric, the effect of angle prevails. Our approach includes an empirical scaling analysis, comparing gravity, curvature-induced force, and viscosity forces on a single bead, yielding a unified empirical viscous force law, and enhancing understanding of Rayleigh-Plateau  regime dynamics. This new framework enriches our understanding of the forces at play in Rayleigh-Plateau instability and provides practical insights into the manipulation of fluid dynamics in industrial applications.
\end{abstract}

\maketitle

\section*{Introduction}

\begin{figure}
\centering
\includegraphics[width=0.85\textwidth]{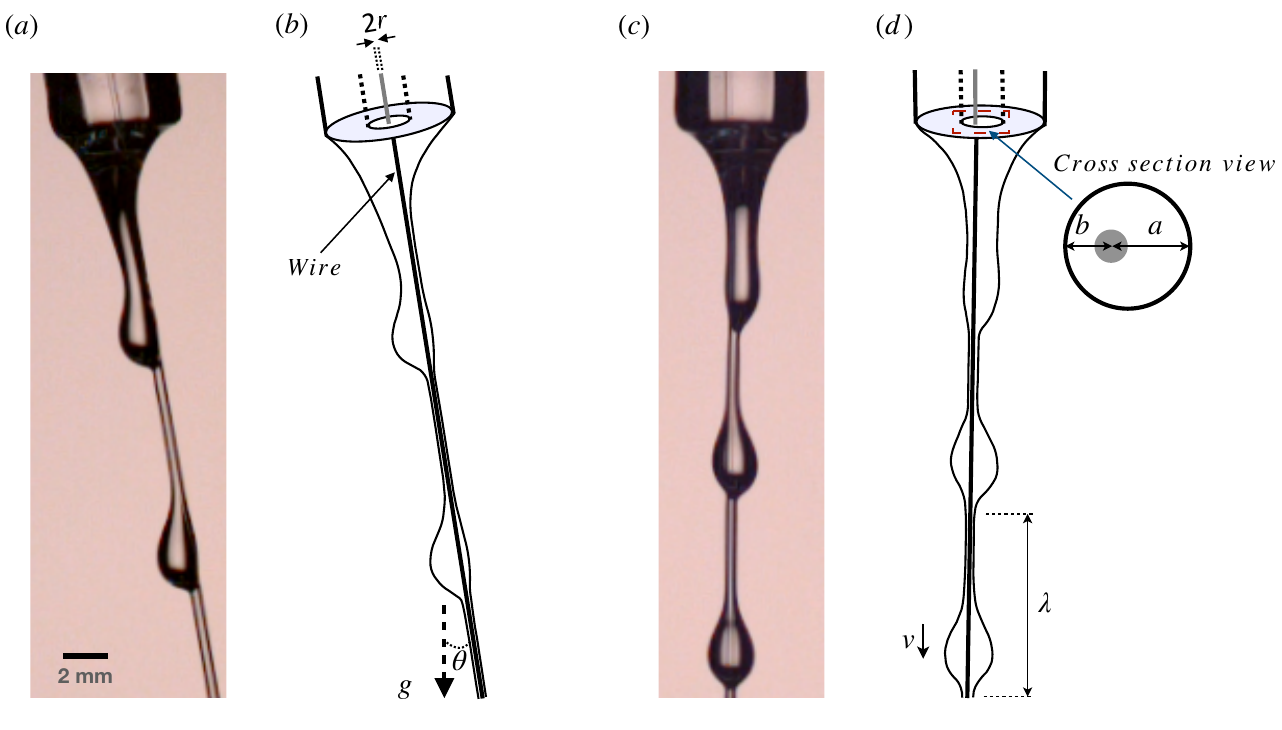}
\caption{\label{fig_schematic} \textbf{Rayleigh-Plateau instability on an angled and eccentric wire.} (a) Experimental image (b) Schematic representation depicting the jet and bead shape near the nozzle at a $10^\circ$ angle ($\theta$) from the vertical with the wire positioned at the center of the nozzle ($e_n=(a-b)/(a+b)=0$). (c) Experimental image (d) Schematic representation illustrating the jet and bead shape near the nozzle at an eccentricity of $e_n=0.23$ for a vertically oriented wire. The diameter of the nylon wire, and nozzle inner diameter is 0.2 mm and 2 mm, respectively.}
\end{figure}

When a wire is placed along the central axis of a nozzle, the liquid flowing through the nozzle gives rise to a hydrodynamic instability \cite{Quere_1990, Kliakhandler_2001}, 
leading to the formation of a string of droplets on the wire due to the Rayleigh-Plateau instability \cite{Plateau_1873,Rayleigh_1892} as shown in Fig.~\ref{fig_schematic}. Typically, three phenomena can occur: i) an isolated regime where primary droplets are separated by secondary droplets (dripping from the nozzle), ii) a periodic Rayleigh-Plateau regime with a constant velocity and bead spacing (jetting from the nozzle), iii) an aperiodic convective regime characterized by random coalescence between primary beads (jetting from the nozzle) \cite{Kliakhandler_2001}. 
 Liquid flowing on a wire is commonly employed for heat and mass exchange between liquids and gasses\cite{Kreith_1988} (e.g., distillation\cite{Sadeghpour_2019}, absorption, fog harvesting\cite{Bai_2011,Ju_2012}, and microfluidics\cite{Gilet_2009}, etc.). 
Liquid flowing down a wire offers the prospect of achieving high heat and/or mass transfer by increasing liquid-side circulation \cite{Nozaki_1998, Wehinger_2013} and decreasing gas-side pressure drop \cite{Hattori_1994, Grunig_2010, Grunig_2012}. Additional benefits include a reduction in dry area \cite{Chinju_2000}, velocity control \cite{Kliakhandler_2001}, and surface area control \cite{Grunig_2010, Grunig_2012, Wagstaff_2023_review}. 
Although the manipulation of the Rayleigh-Plateau instability on a wire stands as a crucial aspect of current research, there is little research into how it behaves when the wires are no longer vertical and/or the wire is no longer co-centric with the liquid nozzle.

The Rayleigh-Plateau instability on a vertical wire depends on the fluid properties\cite{Smolka_2008}, fluid rheology\cite{Boulogne_2012,Gabbard_2023}, wire diameter\cite{Duprat_2007, Duprat_2009, Gabbard_2021}, wire geometry\cite{Xie_2021}, wettability\cite{Haefner_2015}, external forcing \cite{Duprat_2009} and nozzle inner/outer diameter\cite{Sadeghpour_2017, Ji_2019, Ji_2020}. Karimi et al. \cite{Karimi_2023, Karimi_2025} were the first to investigate the effect of inclination angle on liquid flow along a wire, focusing on dripping, jetting, and drop-off phenomena in thin film flows. Similarly, our study examines angled wires and also observed the drop-off phenomenon. However, unlike their work, we focused on the Rayleigh-Plateau instability and the associated bead properties and scaling relationships.
Eghbali et al. \cite{Eghbali_2022} studied the effect of eccentricity exclusively for vertical configurations. In contrast, our research extends this analysis by exploring the effect of eccentricity even when the wire is inclined. Also, Dejean et al. \cite{Dejean_2020} designed a tetrahedron wire packing and observed unique flow behavior related to the angle between legs, and Chinju et al.\cite {Chinju_2000} commented that eccentricity could affect droplet spacing. This is interesting because in an industrial setting the wires could easily be angled or eccentric from either design or manufacturing. Furthermore, this approach holds great utility where adjusting the primary parameters are constrained. For instance, if  the flow rate, wire diameter, wettability, and wire geometry are fixed, one could still angle the wire to control the Rayleigh-Plateau instability based on our findings. 

It is well accepted that the Rayleigh-Plateau  regime shifts to the convective regime with increasing flow rates when all other parameters remain the same\cite{Kliakhandler_2001}. The phenomenon requires that the fluid be sufficiently viscous and the four main forces in competition are gravity, inertia,  surface tension, and viscosity. Theoretical investigations\cite{Kliakhandler_2001, Craster_2006, Duprat_2007, Smolka_2008, Duprat_2009, Ji_2019} begin by estimating Nusselt thickness. Nusselt thickness is defined by the undisturbed film thickness provides a base state, derived by neglecting the surface tension and inertia and determining a velocity profile on the thickness. The undisturbed jet is assumed to be a cylinder because of analytical constraints, yet, in most experiments the jet is conical. Therefore, the theory often doesn't capture the instability quantitatively. 
Here, we derive empirical laws for the dominant forces by performing a force balance on a single droplet specifically within the Rayleigh-Plateau regime, below the pre-instability region. While this approach is more effective in quantifying the governing forces, which were difficult to predict in previous theoretical studies, it is also unable to capture the regime transition.


The current investigation involves a deliberate alteration of the Rayleigh-Plateau instability through two main adjustments: (i) manipulating the wire angle, and ii) varying the eccentricity.
    We also present a scaling analysis within the Rayleigh-Plateau  regime that juxtaposes the impacts of gravity, surface tension force, and viscous forces. 



\begin{figure*}
\centering
\includegraphics[width=0.9\textwidth]{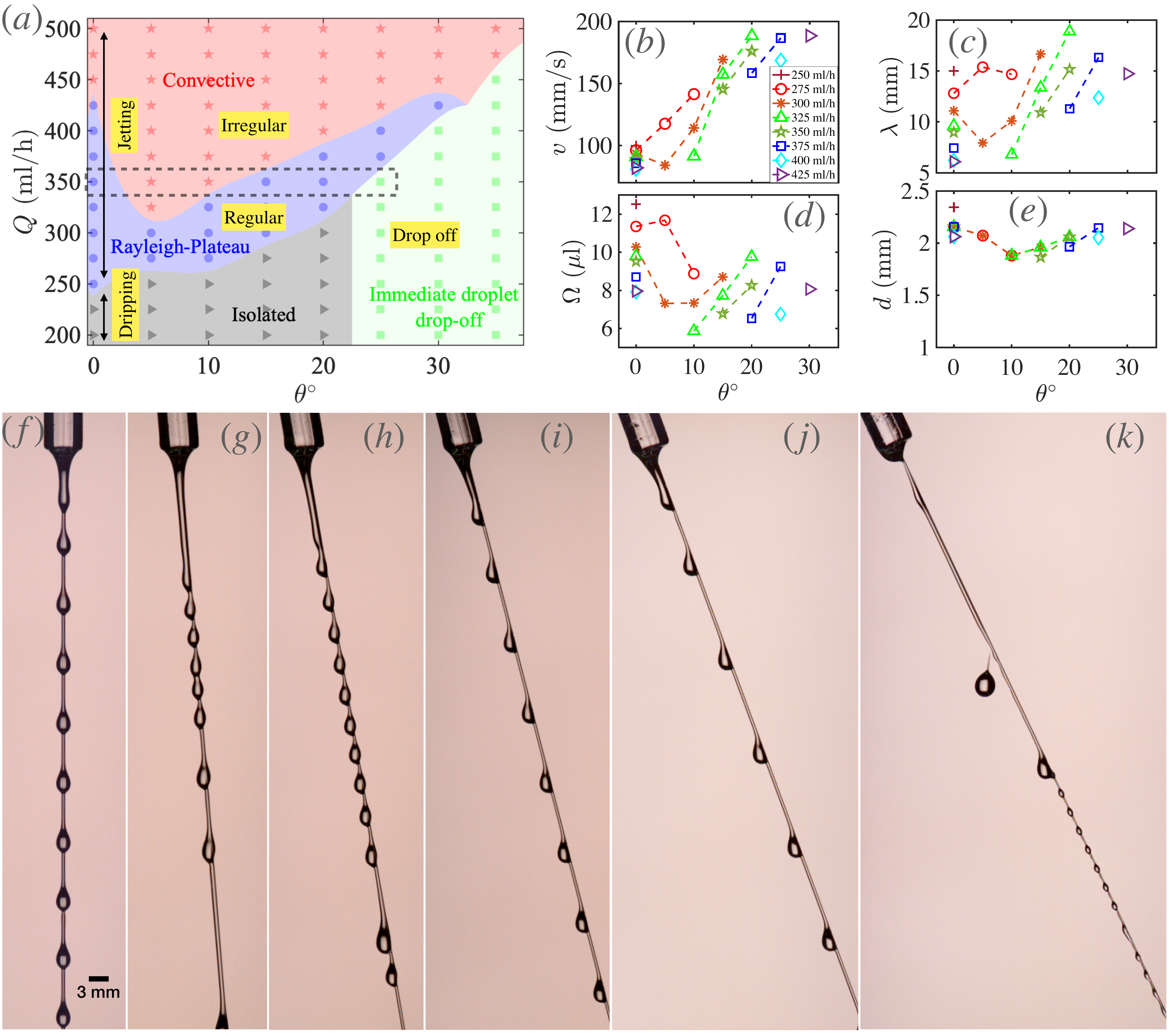}
\caption{\label{fig_angle} \textbf{Influence of angle on the Rayleigh-Plateau instability.} (a) Map illustrating different beading regimes within the parameter space of flow rate and angle of orientation for wire radius $r$ = 0.1 mm and 50 cSt silicone oil. Shaded regions approximate boundaries, and data points highlight domains where isolated, Rayleigh-Plateau, convective, and immediate droplet drop-off events are observed. Variation of bead ($b$) velocity, ($c$) wavelength, ($d$)  volume, and ($e$)  diameter with angle of orientation for the data of the Rayleigh-Plateau  regimes for various flow rates. Disconnected dashed lines in $b-e$ indicate the absence of Rayleigh-Plateau  between data points. Images from cases marked by the rectangular dashed box in ($a$) depicting beads at ($f$) $\theta=0^\circ$, ($g$) $\theta=5^\circ$, ($h$) $\theta=10^\circ$, ($i$) $\theta=15^\circ$, ($j$) $\theta=20^\circ$, and ($k$) $\theta=25^\circ$.}
\end{figure*}

\section{Experimental Methods}


In our experimental setup, we employ silicone oil with a density of 963 kg/m$^3$, kinematic viscosity of $50$ cSt and 100 cSt, and surface tension of 20.8 mN/m at $22^\circ$ C. This silicone oil flows down an angled/eccentric nylon wire, with four nylon wires of radii $r$=0.1, 0.15, 0.2 and 0.25 mm. Fig.~\ref{fig_schematic} and supplementary Fig S1 illustrate the experimental configuration. In the first configuration of this study, we rotated the whole experimental configuration to understand the effect of the angle on the Rayleigh-Plateau instability. We rotate the setup from the vertical at an angle of $\theta$ that varies from $0^\circ$ - $40^\circ$ at intervals of $5^\circ$ as above around $40^\circ$ no jet follows the wire. The setup includes a cylindrical container connected to a nozzle, continuously filled by a syringe pump to maintain a constant flow rate ($Q$). The fluid height in the cylindrical container initially rises and eventually stabilizes at a fixed level when the inflow and outflow rates become equal. Measurements were taken after this stabilization period and we vary the flow rate from 100 ml/h to 500 ml/h at interval of 25 ml/h. The nozzle has inner and outer diameters of 2 mm and 6 mm, respectively. A weight is attached to ensure a stretched wire condition, 
Precision control of the wire angle and eccentricity is maintained using X-Y stages. Bead dynamics are recorded at 1500 fps using a Phantom T3610 camera, and an LED backlight illuminates the bead-wire system. 
Each experiment was observed for a few minutes to accurately characterize the distinctive regimes. 
Note, experiments were repeated three times for each data point in all figures, and resulting error bars are smaller than the marker size.
Bead dynamics were recorded at a resolution of 1280$\times$800 pixels, with droplet velocity determined using TRACKER software\cite{Douglas_2016}. MATLAB is used to determine various properties, including bead spacing (wavelength), volume, and bead size. From the images captured by the camera, the bead volume is calculated by approximating the droplet as a series of small cross-sectional disks with a thickness equal to the pixel size. Numerical integration is then performed across the entire bead to compute its total volume, accounting for and subtracting the volume of the wire (for detailed information, refer to supplementary section II). 

\section{Results}\label{sec2}


\subsection{The effect of wire angle on the Rayleigh-Plateau instability}
Our first objective is to scrutinize the variations in wire orientation to manipulate the Rayleigh-Plateau instability without altering any other parameters. In our experiment, we found similar regimes to Kliakhandler et al.\cite{Kliakhandler_2001} who classified the beading patterns resulting from thin film flow on a wire into three main categories: isolated regime, Rayleigh-Plateau regime, and convective regime. The primary focus of our study is to examine the transition boundary from Rayleigh-Plateau regime to convective regime and gain insights into bead properties within the Rayleigh-Plateau  regime. Such insights are crucial for enhancing efficiency in practical applications \cite{Zeng_2018}.

When the wire is placed vertically (i.e., $\theta=0^\circ$) the Rayleigh-Plateau  regime is observed within the flow rate range of $250-425$ ml/h for wire radius $r$ = 0.1 mm and 50 cSt silicone oil as shown in Fig.~\ref{fig_angle}$a$, with isolated and convective regimes below and above this range of flow rates, respectively. As the angle of the wire is increased the required flow rate for the Rayleigh-Plateau  regime increases and the range of flow rates is reduced.  
The isolated regime switches to immediate droplet drop-off when $\theta>20^\circ$ as the weight of the drop overcomes surface tension forces that maintain the drop on the wire\cite{Karimi_2023, Karimi_2025}.  
Convective regime cases also lead to droplet drop-off at larger $\theta$ but not immediately, also observed in \cite{Karimi_2023, Karimi_2025}. This happens as an interaction of irregular droplets at a larger flow rate and larger angle. An intriguing observation is made where the Rayleigh-Plateau  regime switches to the convective regime and back again when considering increasing angles at flow rates between $325-425$ ml/h. For example, images taken from the dotted rectangular region of Fig.~\ref{fig_angle} 
revealing the transition from Rayleigh-Plateau  to convective regime ($f$ to $g$), convective to Rayleigh-Plateau  regime ($h$ to $i$), and Rayleigh-Plateau  regime to immediate drop-off happens ($j$ to $k$); (see supplementary videos 1-6 for the dynamics \cite{supple_video_2024}).

At $\theta = 0$ a symmetric undisturbed jet (pre-instability jet)  is converted to a series of symmetric droplets as shown in Fig.~\ref{fig_angle}$f$. 
The spacing between successive beads, denoted as wavelength ($\lambda$) decreases with increasing flow rate while the downward bead velocity along the wire ($v$) also decreases with flow rate but less prominently, eventually increasing flow rate transforms into the convective regime.  This transition is typically associated with experimental conditions where the bead spacing and the droplet velocity decrease.  When the wire is angled Fig.~\ref{fig_angle}$i$ \& $j$ show that the symmetry of the jet and droplets are disrupted. This asymmetry results in changes in the bead properties we can measure in the Rayleigh-Plateau  regime. At $Q=$ 300 ml/h the bead velocity ($v$) along the wire exhibits a distinctive pattern of initial decrease followed by an increase for a specific flow rate ($Q$), eventually transitioning into the isolated regime or immediate droplet drop-off, as depicted in Fig.~\ref{fig_angle}$b$. In Fig.~\ref{fig_angle}$c$, the wavelength ($\lambda$), undergoes a similar behavior for the same flow rate, decreasing and then increasing up to the onset of the the isolated regime or immediate droplet drop-off.
In the vertical configuration at $Q = 350$ ml/h, the Rayleigh-Plateau  regime is observed. As the angle increases to $\theta = 5^\circ$ and $10^\circ$, the Rayleigh-Plateau  regime transitions to the convective regime but reappears at $\theta = 15^\circ$ and $20^\circ$. Fig.~\ref{fig_angle} (g and h) and the corresponding supplementary movie (1–5) for 350 ml/h reveal that bead spacing and velocity are smaller near the pre-instability region (even within the convective regime) compared to Fig.~\ref{fig_angle} (f and i), which correspond to the Rayleigh-Plateau regime. This again confirms that the regime shift occurs when both wavelength and velocity drop below a certain threshold for a given flow rate. Furthermore, the initial decrease followed by an increase in wavelength and velocity with the angle appears to be a general trend for a given flow rate (see supplementary section III). However, for certain flow rates, this trend is not observed, likely due to the transition from the Rayleigh-Plateau  regime to the convective regime occurring mid-way through the angle range.
The fluid volume ($\Omega$) within one wavelength (see Fig.~\ref{fig_schematic} for the portion of one single wavelength) exhibits analogous behaviors concerning angle and flow rate, resembling the trends observed in $v$ and $\lambda$, as illustrated in Fig.~\ref{fig_angle}$d$. The bead diameter remains nearly constant Fig.~\ref{fig_angle}$e$.  

\subsection{The effect of eccentricity on the Rayleigh-Plateau instability in vertical orientation}

For the second experimental configuration, we implemented a vertical wire setup, manipulating the eccentricity of the wire at the nozzle to explore its influence on the Rayleigh-Plateau instability under subtle adjustments, as depicted in Fig.~\ref{fig_schematic}c. 
Nozzle eccentricity is quantified as $e_n=(a-b)/(a+b)$, where $a$ and $b$ represent the distances of the wire from the nozzle wall to the center of the wire on the right and left sides of the wire, respectively.  

\begin{figure*}
\centering
\includegraphics[width=0.9\textwidth]{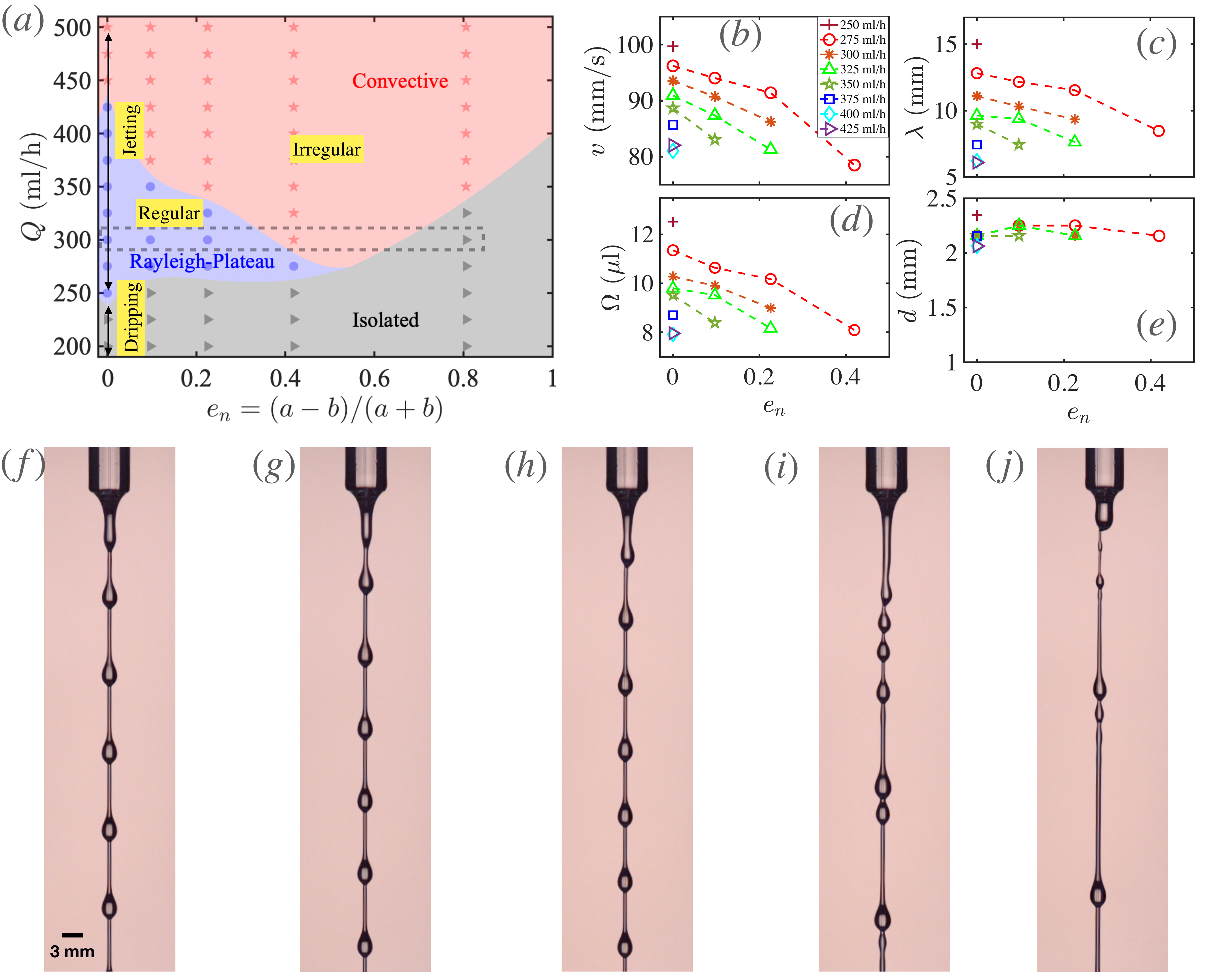}
\caption{\label{fig_eccentric_0} \textbf{Influence of eccentricity on the Rayleigh-Plateau instability for a vertically oriented wire.} ($a$) Regime map within flow rate and eccentricity at $\theta=0^\circ$ for wire radius $r$ = 0.1 mm and 50 cSt silicone oil. Shaded colors depict the approximate regions, while discrete data points indicate the domains where the isolated, Rayleigh-Plateau, and convective regimes are observed. Variation of bead ($b$) velocity, ($c$) wavelength, ($d$) volume, and ($e$) diameter with eccentricity for different flow rates in the Rayleigh-Plateau regime, v.  Images of the beads at $Q=300$ ml/h for $e_n$: ($f$) $0$, ($g$) $0.1$, ($h$) $0.23$, ($i$) $0.42$, and ($j$) $0.81$.}
\end{figure*}

This configuration also yields three regimes similar to those identified by Kliakhandler et al.\cite{Kliakhandler_2001}. Fig.~\ref{fig_eccentric_0}a illustrates a regime map in the parameter space of $Q$ and nozzle eccentricity ($e_n$) for $r$ = 0.1 mm and 50 cSt silicone oil. This configuration enables us to comprehend the transitions between these regimes and their impact on bead properties. Specifically, when the wire is vertically oriented ($\theta=0^\circ$), we observe Rayleigh-Plateau  regimes within the flow rate range of $250-425$ ml/h in Fig.~\ref{fig_eccentric_0}a. Isolated regimes and convective regimes are observed below and above this flow rate range, respectively. The Rayleigh-Plateau  regime diminishes and disappears up to $e_n=0.42$, beyond which only the transition from isolated to convective regime is observed. Experimental images at $Q=300$ ml/h depicting the three regimes are presented in Fig.~\ref{fig_eccentric_0}$f-j$ (see supplementary videos 7-11 for the dynamics \cite{supple_video_2024}). When the wire is eccentric ($e_n\neq0$), the pre-instability jet and the droplets just below the instability position exhibit asymmetry, but droplets below that level become symmetric (see Fig.~\ref{fig_eccentric_0}$h$ for most obvious Rayleigh-Plateau  asymmetry near the nozzle). Both the asymmetrical undisturbed jet and transition from asymmetric to symmetric droplets arise due to the eccentricity of the wire alters the magnitude of the acting upward viscous force, leading to significant changes in the observed regimes and bead properties.

Focusing again on the bead properties within the Rayleigh-Plateau  regimes we get straightforward relationships. In Figs~\ref{fig_eccentric_0}$b-d$, velocity, wavelength, and volume, all decrease monotonically for increasing $e_n$ and $Q$. 
In addition, droplet diameter remains nearly constant as shown in Fig.~\ref{fig_eccentric_0}$e$.  

\begin{figure*}
\centering
\includegraphics[width=0.9\textwidth]{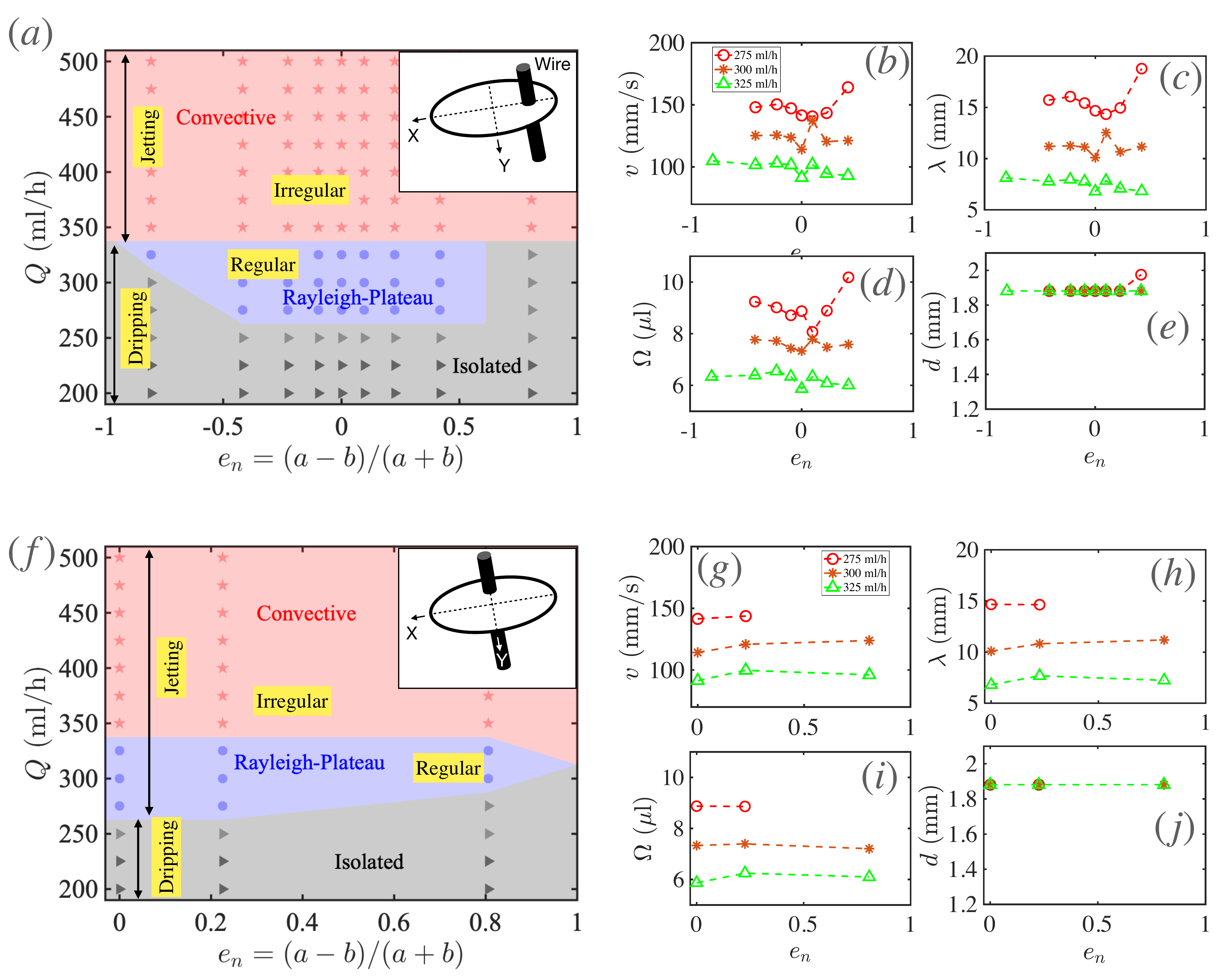}
\caption{\label{fig_eccentric_10} \textbf{Influence of eccentricity on the Rayleigh-Plateau instability for an angled wire.} Regime map illustrating the beading regimes in the parameter space of flow rate and eccentricity at $\theta=10^\circ$ for the shift of ($a$) $x$ direction ($f$) $y$ direction for wire radius $r$ = 0.1 mm and 50 cSt silicone oil. Shaded colors depict the approximate regions, while discrete data points indicate the domains where the isolated, Rayleigh-Plateau , and convective regime are observed. Variation of ($b,g$) velocity, ($c,h$) wavelength, ($d,i$) volume, and ($e,j$) diameter with the nozzle eccentricity ($e_n$) for different flow rates ($x$ direction, $y$ direction) in the Rayleigh-Plateau regime}. 
\end{figure*}

\subsection{The effect of eccentricity on the Rayleigh-Plateau instability in  angled orientation}

We continue by combining angle and eccentricity. The eccentricity can take on two directions within the angled nozzle due to the disruption in symmetry. These directions are designated as $x$ and $y$ at the nozzle-tip tilted reference frame with respect to gravity ($g$) as shown in the insets of Figs~\ref{fig_eccentric_10}$a \& f$. 
The eccentricity is adjusted from $-1$ to $1$ in the $x$ direction and $0$ to $1$ in the $y$ direction due to symmetry. 
Figs~\ref{fig_eccentric_10}$a \& f$ display a regime map for the $x$ and $y$ eccentricities at $\theta=10^\circ$ for $r$ = 0.1 mm and 50 cSt silicone oil and show that the isolated and Rayleigh-Plateau  regimes are affected only at the extremes. 
The velocity, wavelength, and volume exhibit mild dependence on eccentricity as depicted in Fig.~\ref{fig_eccentric_10}($b, c, d, g, h, \& i$). Note that when $\theta$ is introduced the effect of eccentricity becomes less significant. This is because gravity dominates the direction of the jet and droplet asymmetry over the eccentricity for higher angles ($\theta\gtrsim10^\circ$) .

\subsection{Scaling law on the Rayleigh-Plateau instability on wire}
In earlier investigations \cite{Kliakhandler_2001, Craster_2006, Duprat_2007, Smolka_2008, Duprat_2009, Ji_2019}, researchers established a transition boundary between Rayleigh-Plateau  and convective regimes. This was accomplished by delineating the Nusselt thickness by focusing solely on the interplay between viscous and gravity forces through the Navier-Stokes equations and solving for velocity to yield the thickness of the film on the wire. Notably, they accomplish a solution to the velocity profile by ignoring surface tension, which is later added to address the instability.  Consequently, the theoretical model implies that an undisturbed jet (pre-instability jet) maintains a purely cylindrical shape, in contrast to the observed conical form in most experiments (Fig.~\ref{fig_eccentric_0}$f$). This discrepancy leads us to question the realism and practicality of these theoretical approaches. 
Here, we propose a novel perspective by examining the interplay of three forces in the Rayleigh-Plateau  regime: gravity, surface tension, and viscosity acting on a single bead that enables us to derive an empirical law for viscous and surface tension forces, which are otherwise highly challenging to address through theoretical methods. However, we acknowledge that our approach is unable to predict the transition.

\begin{figure*}
\includegraphics[width=0.95\textwidth]{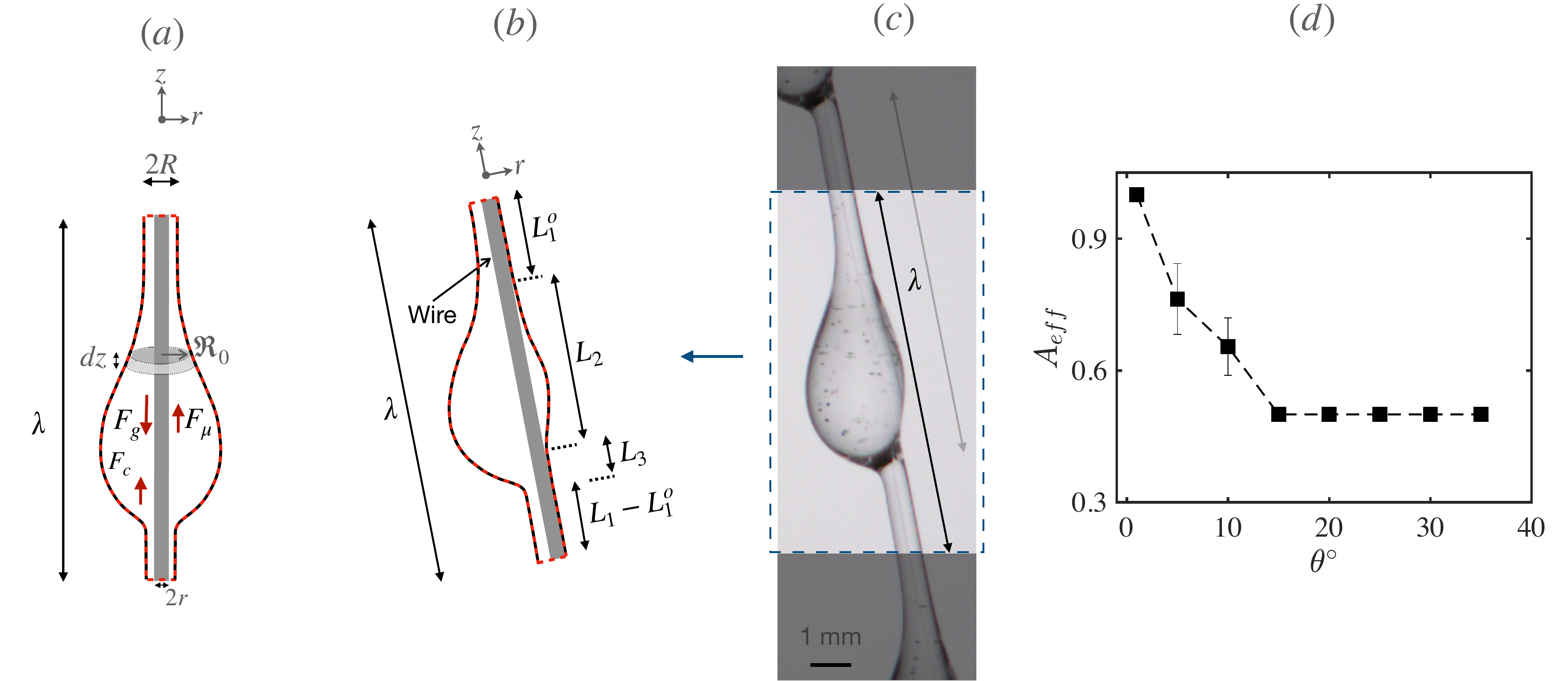}
\caption{\label{fig5_viscous} {\textbf{Viscous force for Rayleigh-Plateau Instability on a wire.} (a) Schematic representation of a bead for $r$ = 0.1 mm, $e_n=0$ and 50 cSt silicone oil at (a) $\theta=0^\circ$, and (b) $\theta=10^\circ$ (c) experimental image of a bead shown in (b).
 (d) Variation of $A_{\text{eff}}$ with $\theta$, including error bars as indicated.}}
\end{figure*}


We adopt a single-bead approach (control volume) to elucidate the empirical law, where the bead comprises two segments: a slender cylindrical part and a substantial droplet section (See Fig.~\ref{fig5_viscous}~$a \&b$). In our model, the control volume is taken over a length corresponding to one wavelength, $\lambda$ as shown in Fig.~\ref{fig5_viscous}~a\&b.
{We can define the forces in the $z-$direction on the control volume as:
\begin{equation}{\label{eq_fb_1}}
\rho \Omega\frac{dv}{dt}= -F_g + F_\mu+F_{c}.
\end{equation}}
{If we only consider droplets within the Rayleigh-Plateau regime where the droplets are evenly spaced with constant velocity then  $dv/dt =0$. Thus, our hypothesis posits that the downward force from gravity ($F_g$) is equal to the combined upward force from viscous dissipation due to wire-fluid interaction ($F_{\mu}$) along the wire axis ($z$) and a pressure imbalance force which arises from the horizontally asymmetric shape of the droplet ($F_{c}$) \cite{backholm2020water, keiser2020universality,daniel2017oleoplaning}.} 

Calculating the force ($F_g$) due to gravity is straightforward: $F_g = \rho \Omega g \cos{\theta}$, with $\rho$ and $g$ denoting liquid density and gravitational acceleration. 


{The viscous force ($F_\mu$) is quantified as proportional to dynamic viscosity, area ($dA$), and the velocity gradient ($dv/dr$)}
{
\begin{equation}
F_{\mu}=\int_{A}{\mu {dA} \frac{dv}{dr}}.\\
\label{eq2_1}
\end{equation}
The force equation is derived by integrating the differential viscous force contributions over a control volume of length $\lambda$, as illustrated in Fig.~\ref{fig5_viscous}a. Here, $\mathfrak{R}_0(z)$ denotes the local radius of the liquid–air interface at axial position $z$. The elementary surface area of the wire over an infinitesimal segment $dz$ is given by $dA = 2\pi r\,dz$, where $r$ is the radius of the wire.
Although the local fluid thickness at position $z$ is $\mathfrak{R}_0(z) - r$, the internal flow visualization using tracer microparticles (see Supplementary movie 12 \cite{supple_video_2024}) suggests that the velocity gradient near the wire develops over a much smaller distance ($r$), confined close to the wire surface. Therefore, the velocity gradient can be taken as $dv/dr \sim v/r$ near the wire.
Following these assumptions, and starting from first principles, Eq.~(1) can be simplified for the vertical configuration as:
\begin{equation}
F_{\mu} \propto \mu \int_{z=0}^{\lambda}2 \pi r \left({\frac{v}{r}}\right) {dz}\\ = T \mu v \lambda,
\label{eq2}
\end{equation}
where $T/2\pi$ is the proportionality constant.} 

As the inclination angle deviates from zero, the bead becomes increasingly asymmetric, resulting in a reduction of the viscous force. Due to this asymmetry integration of Equation~\ref{eq2_1} is analytically intractable from first principles. To address this complexity, we introduce an effective parameter, $A_{\text{eff}}$, defined phenomenologically as a function of the inclination angle to capture the influence of asymmetry on the viscous force. We assume that the viscous force scales with the contact area between the droplet and the wire; however, under inclination, the asymmetric shape and motion of the droplet effectively reduce this area.
This empirical derivation mirrors Stokes' law\cite{Stokes_1851}, a frictional force model for spherical objects in a low Reynolds number implying a high viscous force setting.
{We experimentally measure $A_{\text{eff}}$ by witnessing the shape of bead as depicted in Fig.~\ref{fig5_viscous}$c$ and supplementary movie 12 \cite{supple_video_2024}, and the exponents of it obtained through fitting.} We notice that the shape of the droplet is no longer symmetric for $\theta>0$. Consequently, We track the interface of a droplet on both sides of the wire (Fig.~\ref{fig5_viscous}$b$). On the \emph{left side} of the wire within the regions of lengths $L_1$, $L_2$ and $L_3$ ($\lambda=L_1+L_2+L_3$), the air-liquid interface is separated from the left wire surface by some amount of fluid. Whereas, on the \emph{right side} of the wire, the air-liquid interface is only separated from the wire along $L_2$. Thus, we assume that viscous drag acts on the left half of the wire of length $\lambda$ and right half of the wire of length $L_2$. Consequently, $A_{\text{eff}}$ is defined as the ratio of the affected area for viscous drag to the total surface area as
\begin{equation}
A_{\text{eff}}=\frac{\pi r L_1+2\pi r L_2+\pi r L_3}{2\pi r \lambda}=\frac{1}{2}(1+\frac{L_2}{\lambda}).
\end{equation}
 For all vertical cases, we observe $L_2=\lambda$ as the shape of the bead is symmetric and the liquid is present on both sides of the wire. For $\theta\geq15^\circ$  we found $L_2=0$ as there is no fluid present on the right side due to the vertical component of gravity concerning the wire axis. Therefore, $A_{\text{eff}}=1$  for all vertical cases and and $A_{\text{eff}}=0.5$ for the cases where $\theta\geq 15^\circ$, and is unaffected by variations in wire diameter and flow rate (Fig.~\ref{fig5_viscous}d) given our experimental conditions. Notably, there is a pronounced variation in $A_{\text{eff}}$ (between 0.5 and 1) observed at lower wire angles $\theta=5^\circ$ $\&$ $10^\circ$ with the specified error bar  as shown in (Fig.~\ref{fig5_viscous}d). {The fitted exponent of $A_{\text{eff}}$ was found to be $2 \pm 0.2$. 
This outcome stems from the fact that the formulation of $A_{\text{eff}}$ is based on a two-dimensional visualization, 
where we approximate that only half of the circumference contributes or not. 
In reality, this contribution is not exactly one-half and cannot be determined precisely, 
as only two-dimensional information is available.
Thus, by combining the best-fit result with this approximate estimate of $A_{\text{eff}}$, 
we arrived at a prediction for the viscous force.}

This is further validated by the observed increase in velocity and wavelength when the system transitions back from the convective regime to the Rayleigh–Plateau regime. 
This transition suggests a force imbalance acting on the droplet. In the convective regime, inertial effects become significant, leading to a net reduction in the upward force on the droplet. The implication is that the Galilei number is important, defined as $Ge = F_g / F_\mu = \rho \Omega g \cos{\theta} / (T \mu v \lambda A^2_{\text{eff}})$, which represents the ratio of gravitational to viscous forces. Here, increasing the flow rate leads to a substantial decrease in the wavelength $\lambda$, while other parameters remain approximately constant. We attribute this reduction to a substantial decrease in the upward viscous force, while the downward gravitational force remains largely unchanged. This results in a significant increase in the Galilei number, indicating a growing dominance of inertial effects in the convective regime.

\begin{figure*}
\includegraphics[width=0.95\textwidth]{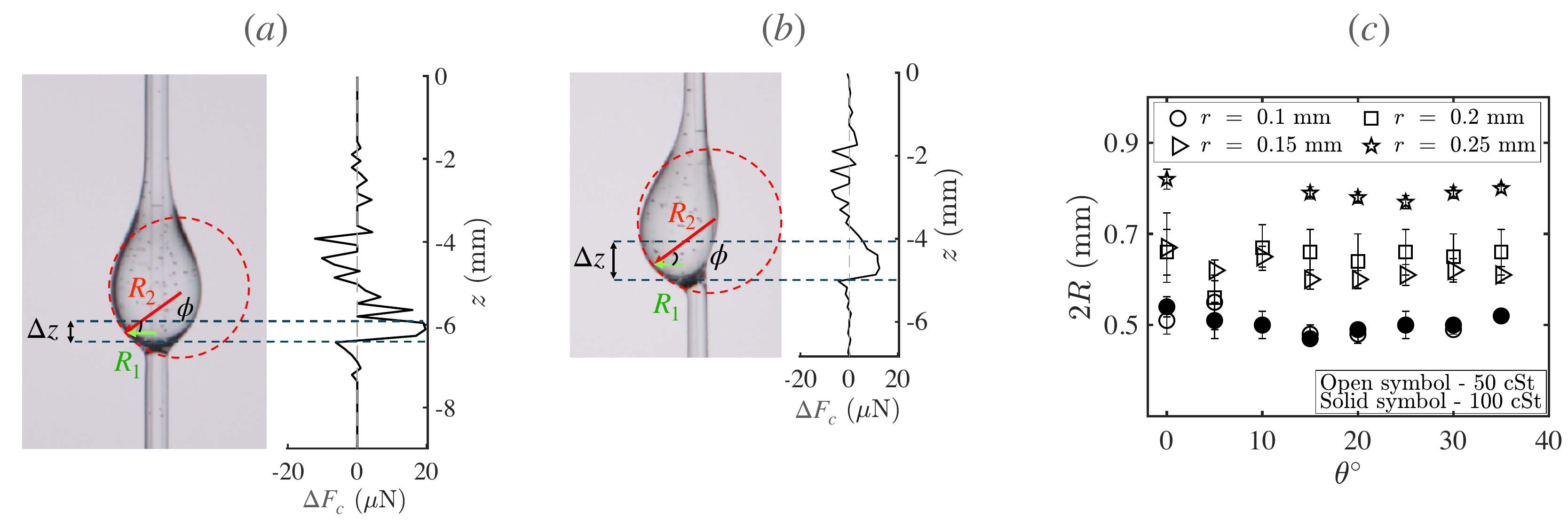}
\caption{\label{fig6_curvature} {\textbf{Curvature-induced force for Rayleigh-Plateau Instability on a wire.} (a) Experiment image of a bead and $z$ dependent upward force for $r$ = 0.1~mm, $e_n=0$ and 50~cSt silicone oil at (a) $\theta=0^\circ$, and (b) $\theta=10^\circ$
 (d) variation of the slender
cylinder diameter (2$R$) is shown as a function of $\theta$ for all cases within the Rayleigh-Plateau regime.}}
\end{figure*}

\begin{figure*}
\includegraphics[width=0.6\textwidth]{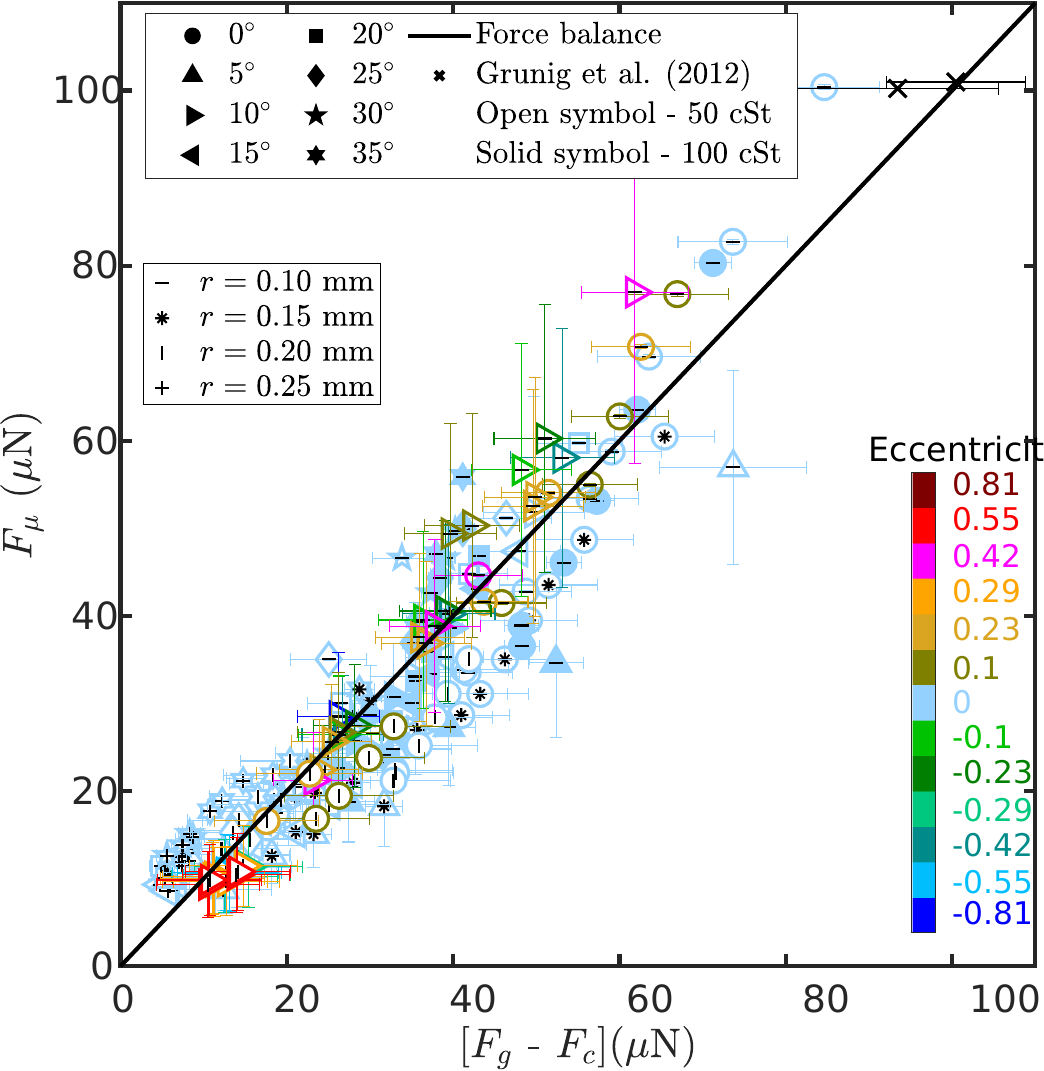}
\caption{\label{fig7_scaling} \textbf{Unified scaling law of Rayleigh-Plateau Instability on a wire.} Unified scaling for the force balance equation (Equation~\ref{eq_fb}) across different wire radii and viscosities in the Rayleigh-Plateau regime.}
\end{figure*}

{
In our experiments, all of the droplets adopt an asymmetric shape, as shown in Fig.~\ref{fig6_curvature}a,b, indicating a broken horizontal symmetry and the emergence of a net curvature force ($F_c$). 
Integration of the Laplace pressure ($p$) over the control volume yields the force
\begin{equation}
F_{c}= \int_{A}p \hat n \cdot \hat z dA = \int_{z=0}^{\lambda}\sigma \left( \frac{1}{R_1(z)} + \frac{1}{R_2(z)} \right) 2\pi R_1(z)  \sin{\phi(z)} dz
\label{eq:Fc_general}
\end{equation}
where $R_1$ and $R_2$ are the principal radii of curvature at  $z$, $\hat n$ is the unit normal vector to the surface, and  $\phi$ is the angle between the surface normal and the horizontal. In Fig.~\ref{fig6_curvature}a,b an image from the experiment is shown and we outline the radius of curvature ($R_2$) at one location and notice that the radii are larger at the top of the droplet than at the bottom. We also outline $R_1$ which is the axial radius at that point.  Next to the image we plot $\Delta  F_c$ vs. $z$ (direct correspondence to the image) to illustrate the variation of $F_c$ due to the asymmetry of the droplet. The data confirms that a force arises from the curvature of the droplet with a larger significance near the bottom of the droplet. We numerically integrate within the control volume to estimate the contribution of the curvature to the forces and find that the values are close to $2\pi R \sigma$ (see supplementary figure S7), although we note that the integration is very sensitive to curvature estimation through image processing ($\approx \pm10~\mu$N). Thus, we now aim to find an analytic approximation.}

 {
We notice that the primary upward force arises over a finite thickness $\Delta z$ from the bottom of the droplet portion, which is of the order of $2R$ in the vertical configuration Fig.~\ref{fig6_curvature}a and $4R$ in the inclined case Fig.~\ref{fig6_curvature}b. In both scenarios, the outer radius satisfies $R_2 \gg R_1$ and $\phi \sim 30^\circ$. }

For the vertical configuration, the surface area is approximately $2\pi R_1 \Delta z$. Therefore, the curvature-induced force can be approximated as
\begin{equation}
F_{c}  \approx \frac{\sigma}{R_1} (2\pi R_1) (\frac{1}{2}) 2R  = 2\pi R \sigma.
\label{eq:Fc_vertical}
\end{equation}

In the inclined configuration, the right interface contributes negligibly to the upward force. Hence, the effective area can be approximated as $\pi R_1 \Delta z$ and  $\Delta z = 4R$ as previously noted. Despite this, the net curvature-induced force still approximates to
\begin{equation}
F_{c}  \approx \frac{\sigma}{R_1} (\pi R_1) (\frac{1}{2}) 4R  = 2\pi R \sigma.
\label{eq:Fc_vertical}
\end{equation}

Thus, the expression $2\pi R \sigma$ provides a robust first-order approximation of the net curvature-induced force driven by asymmetry.

The diameter of the thin cylinder ($2R$) expands with the wire diameter, yet remains nearly unaffected by the flow rate, as illustrated in Fig.~\ref{fig6_curvature}$c$. Consequently, the force balance equation (Equation~\ref{eq_fb_1}) for the Rayleigh-Plateau  regime can be formulated as

\begin{equation}{\label{eq_fb}}
 \rho \Omega g \cos{\theta} =T \mu v \lambda A^2_{eff}+2\pi R \sigma.
\end{equation}

In Fig.\ref{fig7_scaling}, we illustrate the force balance relationship (i.e., $F_\mu$ vs [$F_g-F_{c}$]) for four wire diameters and two different viscous fluids. 
The force balance equation (Eq.~\ref{eq_fb}) is indicated by the black line with $T=1.4$. All data conform to the force balance equation and are even confirmed by analyzing the Gr\"{u}nig et al\cite{Grunig_2012} data where surface tension and viscosity of the fluids are different than current fluids. Unfortunately, the model is limited to the Rayleigh-Plateau regime, as it relies on constant values of 
$\lambda$ and $v$ to derive empirical laws. In convective regimes, these parameters are not constant, and in the isolated regime, the entire wire is not wetted due to the presence of dry regions.  

\section{Conclusion}
In conclusion, our study has successfully demonstrated the significant impact of manipulating the Rayleigh-Plateau instability by altering both the angle of the nozzle-wire system and the eccentricity of the wire within the nozzle. The investigation revealed a strong dependence of the Rayleigh-Plateau instability regimes on angle and eccentricity. In general, angling the wire decreases the range and increases the average of the range of volumetric flow rate. Eccentricity affects mainly vertical wires where the Rayleigh-Plateau  regime shrinks with increasing eccentricity. The effect of angle dominates over eccentricity when the wire is both angled and eccentric. Within the Rayleigh-Plateau regime noteworthy changes occur in velocity, wavelength, and bead volume. 

Furthermore, to provide a comprehensive understanding, we conducted an empirical scaling analysis, comparing the forces associated with gravity, curvature-induced force, and viscosity on a single bead in the Rayleigh-Plateau regime. The outcome led to the formulation of a unified empirical viscous force law, which serves as a valuable contribution to the understanding of the intricate dynamics within only the Rayleigh-Plateau  regime. 



 \section*{Acknowledgements}
We wish to thank Professor William L. Roberts for the initial idea and inquiry, and Professor Sigurdur Thoroddsen and his laboratory for letting us use their equipment. We also thank Atefeh Pour Karimi, Benoit Scheid, Spencer Truman, Abhijit Kumar Kushwaha, Nilamani Sahoo, John Bush,  Jacco Snoeijer, and the referees for their fruitful discussions. All authors wish to thank KAUST for funding.

\bibliographystyle{apsrev4-2}
\bibliography{prfluidbib.bib}

\end{document}